\documentstyle[12pt]{article}

\def\lromn#1{\uppercase\expandafter{\romannumeral#1}}

\begin{document}
\begin{flushright}
TU/97/524\\
RCNS-97-02\\
\end{flushright}

\vspace{12pt}

\begin{center}
\begin{large}

\bf{
Time Evolution of Unstable Particle Decay Seen with Finite Resolution
}

\end{large}

\vspace{36pt}
\begin{large}
I. Joichi, Sh. Matsumoto, and M. Yoshimura

Department of Physics, Tohoku University\\
Sendai 980-77 Japan\\
\end{large}

\end{center}

\begin{center}
\vspace{54pt}

{\bf ABSTRACT}
\end{center}

Time evolution of the decay process of unstable particles is investigated
in field theory models.
We first formulate how to renormalize the non-decay amplitude beyond
perturbation theory and then discuss short-time behavior of very long-lived
particles.
Two different formalisms, one that does and one that does not, assume
existence of the asymptotic field of unstable particles are considered.
The non-decay amplitude is then calculated 
by introducing a finite time resolution of measurement,
which makes it possible to discuss both renormalizable and non-renormalizable
decay interaction including the nucleon decay.
In ordinary circumstances the onset of the exponential decay law 
starts at times as early as at roughly
the resolution time, but with an enhanced amplitude which
may be measurable.
It is confirmed that the short-time formula $1 - \Gamma t$ of the exponential
decay law may be used to set limits on the nucleon decay rate in
underground experiments.
On the other hand, 
an exceptional example of S-wave decay of very small Q-value
is found, which does not have the exponential period at all.

\newpage

1. {\bf Introduction} 

It is an old problem since the classic works of Dirac, Weisskopf and Wigner
how to describe the decay process of unstable state,
following the principles of quantum mechanics.
It has been repeatedly questioned how the exponential decay law
is modified both at late and early times.\cite{non-exponential decay law} 
New motivation for investigating this age-old problem has been added 
when relevance of the non-observation of the nucleon decay is addressed
\cite{chiu-misra-sudar 82}, \cite{maiani et al 93} 
and whenever high resolution experiments with new technology
are performed to search for deviation from 
the exponential law\cite{test of e-decay: nuclear}, 
\cite{test of e-decay: particle}, \cite{test of e-decay: atomic}.
A conceptual problem, called the quantum Zeno paradox\cite{chiu-sudarshan},
also renewed
interest in this problem, the central issue here being whether
the initial decrease of non-decay probability proceeds via the square of
time, $t^{2}$, which appears inevitable in quantum mechanics, but
not in accord with the exponential law.

A field theoretical description of unstable particle decay has
complication due to renormalization.
It was pointed out in ref \cite{maiani et al 93} 
that the usual mass, coupling and wave
function renormalization is not sufficient to remove all infinities that
appear in the non-decay amplitude at finite times, if one works 
in perturbation theory.
We shall discuss below how this problem may be resolved.

In the present work we first discuss an exact formulation of time
evolution of unstable particle decay without assuming existence
of the asymptotic field of unstable particle.
We then show how renormalization is 
performed beyond perturbation theory, both in this formalism and
in a more conventional approach that assumes existence 
of the asymptotic field.

We next investigate how time evolution of the decay proceeds
when viewed with finite time resolution.
Finite time resolution can be regarded as an effective
means of introducing a resolution in actual observation.
In all applications of practical interest 
the resolution time $\Delta t$ is limited from both below and above, 
$\frac{1}{\Gamma } > \Delta t > \frac{1}{M}$ where $M$ is a physical energy
scale of the process such as the mass of the decaying particle and
$\Gamma $ is the decay rate.
If the time scale one wants to probe into is larger than $\frac{1}{\Gamma }$,
description of the decay as a function of time effectively loses
its meaning, and one looks into some other measurable quantity such
as the energy peak of produced decay products directly.
If the time scale is smaller than $\frac{1}{M}$, 
the concept of particle picture is doubtful.
It is thus sufficient to be able to compute the time evolution of
the non-decay amplitude only for a coarse grained resolution in the time
range above.

The decay law then differs in fine detail, depending on the amount of
resolution.
In ordinary circumstances 
the exponential law becomes excellent  at times as early as of
order $\Delta t$.
The absolute value of the non-decay amplitude is however
enhanced by a resolution dependent amount,
\( \:
\approx 1 + \frac{2}{\pi }\,\Gamma \Delta t 
\: \)
for
\( \:
\frac{1}{2Q} < \Delta t \ll \frac{1}{\Gamma } \,, 
\: \)
and
\( \:
\approx 1 + \frac{\Gamma }{2\pi Q}
\: \)
for $\Delta t \ll \frac{1}{2Q}$, where $Q$ is the $Q$ value of
the threshold.
The short-time behavior of the probability at $t < \Delta t$ shows
a typical time dependence of
\( \:
1 - O[\frac{\Gamma }{\Delta t}]\,t^{2} \,, 
\: \)
characteristic of the quantum mechanical law.
When $\Delta t $ becomes large and approaches the lifetime 
$\frac{1}{\Gamma }$, time evolution shows a large deviation from
the exponential law.

An interesting exception on observability of deviation from
the exponential law occurs when $Q < \Gamma $, which
becomes possible in the $Q \rightarrow 0$ limit of the S-wave decay;
\( \:
\Gamma \approx ({\rm coupling})\; \times \sqrt{MQ} \,.
\: \)
The exponential period may not exist at all in this case.
Whether this case is realized in nature is yet to be found.

Another common view of the unstable particle decay is to consider as 
inseparable the formation process of unstable state 
created from stable many particle states.\cite{resonance idea}
This view introduces a new time scale of formation time interval.
But if the time resolution $\Delta t$ we consider here is much larger than this
formation time interval, then effects of finite time resolution dominate
over the formation effect.
This inequality of two time scales usually holds in most practical cases.
This is fortunate because it is usually difficult to know the 
formation time interval.

A field theoretical approach discussed in detail in Appendix makes it
possible to treat the region of theoretical interest, $t \ll 1/M$.
The infinitely good time resolution is realized in this formalism;
$\Delta t = 0$ formally.
It is found that after renormalization of the proper self-energy 
the non-decay probability has some peculiar behavior in the 
$t \rightarrow 0$ limit, at least in perturbation theory.
Moreover, the limiting behavior is not universal.
In a renormalizable case examined there is a linear rise $\propto 1/t$
in the time range,
\( \:
\frac{\Gamma }{2\pi M^{2}} > t \gg e^{-\,4\pi^{2} /g^{2}}\frac{1}{M}
\,.
\: \)
On the other hand, one has the linear decrease of 
$1 - O[t]$ in a super-renormalizable model.
These results cannot be extrapolated to the true $t\rightarrow 0$ limit,
because one must go into a non-perturbative region.

It thus seems that one cannot answer 
the quantum Zeno paradox of unstable particle
decay in field theory that assumes the asymptotic field.
The true $t \rightarrow 0$ limit, or $\Delta t \rightarrow 0$ limit,
is difficult to deal with in perturbation theory, the only scheme 
that one can compute to a high precision, 
because in approaching $t \rightarrow 0$
one has to know the behavior of Green's function in ever increasing
high energy.

\vspace{0.5cm} 
2. {\bf Projector formalism and decay law in quantum mechanics} 

We shall use an exact integro-differential equation for the non-decay
amplitude,
\( \:
\langle 1|e^{-iHt}|1 \rangle \,, 
\: \)
with $|1\rangle $ the inital state of unstable particle.
Following Peres\cite{peres},
we introduce the projector onto the initial unstable state,
\( \:
P \equiv |1\rangle \langle 1| \,, 
\: \)
and decompose the total Hamiltonian $H$ as
\begin{eqnarray}
&& 
H_{0} \equiv PHP + (1 - P)H(1 - P) \,, \hspace{0.5cm} 
V \equiv H - H_{0} \,. 
\end{eqnarray}
Define energy eigenstates $|a\rangle $ of $H_{0}$;
\( \:
H_{0}|a\rangle = E_{a}\,|a\rangle \,.
\: \)
It is then easy to confirm that the
decay interaction $V$ operates only between
the initial state $|1\rangle $ and its orthogonal compliment $|n\rangle $;
\( \:
V_{1n} = V_{n1}^{*} \neq 0 \,, \hspace{0.5cm} 
V_{mn} = V_{11} = 0 \,.
\: \)
We use the intermediate Roman letters such as $m\,, n$
to denote eigenstates projected by $1 - P$.
The interaction $V$ is defined here in a novel way;
it depends on the prepared initial state.
Although this looks odd, it makes possible some exact analysis, as will
be made clear shortly.

We work in the interaction picture and 
expand the state at a finite time $t$, using the basis of the
eigenstate of $H_{0} \,;$
\( \:
|\psi \rangle _{I} =  \sum_{a}\,c_{a}(t)\,|a\rangle \,.
\: \)
We thus write the time evolution equation for the coefficient $c_{a}(t)$;
\begin{eqnarray}
&&
i\dot{c}_{m} = V_{m1}\,e^{i(E_{m} - E_{1})t}\,c_{1} \,, 
\hspace{0.5cm} 
i\dot{c}_{1} = \sum_{n}\,V_{1n}\,e^{-\,i(E_{n} - E_{1})t}\,c_{n} \,.
\end{eqnarray}
The non-decay amplitude is related to this coefficient by
\begin{equation}
\langle 1|e^{-iHt}|1 \rangle = e^{-iE_{1}t}\,\langle 1|e^{iH_{0}t}e^{-iHt}
|1 \rangle = e^{-iE_{1}t}\,\langle 1|\psi  \rangle_{I} = 
e^{-iE_{1}t}c_{1}(t) \,.
\end{equation}
A closed form of equation for the non-decay amplitude
\( \:
c_{1}(t) \equiv c(t)
\: \)
then follows;
\begin{eqnarray}
&&
\dot{c}(t) = -\,\int_{0}^{t}\,dt'\,\beta (t - t')\,c(t') \,, 
\label{exact int-diff eq} 
\\ &&
\beta (t - t') = \langle 1|V_{I}(t)\,V_{I}(t')|1\rangle =
\int_{\omega _{c}}^{\infty }\,d\omega \,\sigma (\omega )\,
e^{-\,i(\omega - E_{1})(t - t')} \,, 
\\ &&
\sigma (\omega ) = \sum_{m}\,\delta (\omega - E_{m})\,|V_{1m}|^{2}
\,.
\end{eqnarray}
Here
\( \:
V_{I}(t) = e^{iH_{0}t}\,V\,e^{-\,iH_{0}t} 
\: \)
is the decay interaction written in the interaction picture and
the spectral function $\sigma (\omega )$ characterizes interaction between the
unstable particle and decay product particles.
The initial condition $c_{m}(0) = 0$ was used to derive the equation for
$c(t)$, and $\omega _{c}$ is the threshold for the state $|n\rangle $.

The familiar golden rule of perturbation theory immediately follows
if one approximates $c(t')$ in 
the right hand side of eq.(\ref{exact int-diff eq}) by unity and
take the infinite time limit;
\( \:
|c(t)|^{2} \approx 1 - 2\pi \sigma (E_{1})\,t \,.
\: \)
Thus the perturbative decay rate $\Gamma $ is $2\pi \sigma (E_{1})$.
We would however like to elucidate the time evolution in finer detail,
not relying on perturbation theory in this manner.

The standard technique to solve this type of integro-differential equation
is to use the Laplace transform, and one finally 
obtains the non-decay amplitude in the form,
\begin{eqnarray}
c(t) &=& \frac{1}{2\pi i}\,\int_{-\infty }^{\infty }\,d\omega \,
\frac{e^{-\,i(\omega - E_{1})t}}{-\,\omega + E_{1} - G(\omega + i0^{+})} 
\\ 
&\equiv& 
\frac{1}{2\pi i}\,\int_{-\infty }^{\infty }\,d\omega \,F(\omega + i0^{+})\,
e^{-i(\omega - E_{1})t} \,, 
\\ 
G(\omega + i0^{+}) &=& \int_{\omega _{c}}^{\infty }\,d\omega' \,
\frac{\sigma (\omega ')}{\omega ' - \omega - i0^{+}} =
i\,\int_{0}^{\infty }\,d\tau \,\beta (\tau )e^{i\tau (\omega - E_{1})
- 0^{+}\tau } 
\,.
\end{eqnarray}
The initial condition
\( \:
c(0) = 1
\: \)
was imposed in this derivation.

The analytic property of the function,
\begin{eqnarray}
F(z) \equiv \frac{1}{- z + E_{1} - G(z)} \,, 
\end{eqnarray}
is evident;
$F(z)$ is analytic except on the real axis with the branch
cut starting from a threshold, $z > \omega _{c}$.
This makes it possible to write the amplitude in a convenient form,
\begin{eqnarray}
&&
c(t) 
= \int_{\omega _{c}}^{\infty }\,d\omega \,H(\omega )e^{-\,i(\omega - E_{1})t}
\,, \label{sol for non-decay} 
\\ &&
H(\omega ) = \frac{F(\omega + i0^{+}) - F(\omega - i0^{+}) }{2\pi i}
= 
\frac{\sigma (\omega )}{(\omega - E_{1} + \Pi (\omega ))^{2}
+ (\pi \sigma (\omega))^{2}} 
\,, \label{spectral-h} 
\end{eqnarray}
where $\Pi (\omega )$ was defined by the boundary value of $G(z)$,
when $z$ approaches the real axis,
\( \:
G(\omega + i0^{+}) = \Pi (\omega ) + i\pi \sigma (\omega ) \,.
\: \)

It is well known that in the cut $\omega $ plane there is a pole in
the second Riemann sheet, near (and below)
the real axis if the decay interaction is weak enough.
The pole location $z$ is determined by
\begin{equation}
z - E_{1} + G_{\lromn 2}(z) = 0 \,, \label{pole location} 
\end{equation}
where the analytic function $G(z)$, hence $F(z)$, 
is extended into the second sheet by
\( \:
G_{\lromn 2}(\omega - i0^{+}) = G_{\lromn 1}(\omega + i0^{+})
\: \)
through the branch cut at $\omega > \omega _{c}$.
Dominance of this pole term is equivalent to taking a Breit-Wigner
form for $H(\omega )$ in eq.(\ref{sol for non-decay}) and neglecting
the threshold value $\omega _{c}$ by the replacement,
\( \:
\omega _{c}  \rightarrow - \,\infty 
\: \).
The equation (\ref{pole location})
gives perturbatively an imaginary part of the pole,
\( \:
-\,i\Gamma /2 = -\,i\pi \sigma (E_{1}) \,.
\: \)
The ideal exponential decay law 
\( \:
|c(t)|^{2} \propto e^{-\,\Gamma t} 
\: \)
then follows, but this approximation fails both at early and late times.

At late times the threshold behavior of $\sigma (\omega )$ taken to be
\( \:
\sigma (\omega ) =  c\,(\omega - \omega _{c})^{\alpha } 
\: \)
dictates the power law
\( \:
|c(t)|^{2} \propto t^{-\,2(\alpha + 1)} \,
\: \)
\cite{goldberger-watson}. 
The transition time $t_{p}$ from the exponential to the power decay may be
estimated by equating the amplitude in the power period,
\begin{equation}
c(t)_{{\rm power}} \approx 
-\,i c\,\frac{\Gamma (\alpha + 1)}{Q^{2}\,t^{\alpha + 1}}\,
e^{-i(\,\omega _{c} - E_{1}\,)t - i\frac{\pi }{2}\alpha }
\,, \label{power law decay} 
\end{equation}
to the exponential $e^{-\,\Gamma t/2}$ to give \cite{jmy-96} 
\begin{equation}
t_{p} \approx \frac{2}{\Gamma }\,\ln \left( \frac{Q^{2}}{c\,
\Gamma (\alpha + 1)\,\Gamma ^{\alpha + 1}}\right) \,.
\end{equation}
Here $\Gamma(z)$ is the Euler's gamma function, and $Q = E_{1} - \omega _{c}$
is the Q-value for the decay.

By using the "elastic" unitarity relation,
\begin{equation}
F(\omega + i0^{+}) - F(\omega - i0^{+}) = 2\pi i\sigma (\omega )\,
|F(\omega + i0^{+})|^{2}
\,, 
\end{equation}
equivalent to eq.(\ref{spectral-h}), one may deform the contour of
integration on the real axis into a sum of two contours,
the one around the pole as shown 
in C$_{0}$ in Fig.1 and the other along C$_{1}$;
\begin{equation}
c(t) = \frac{1}{2\pi i}\,\left( \,\int_{C_{0}} + \int_{C_{1}}\,\,\right)
\,dz\,F(z)e^{-i(z - E_{1})t} \,.
\end{equation}
The limiting power law formula, eq.(\ref{power law decay}), is a result
of an approximation to the contour integral along $C_{1}$,
which can be used in the large time limit. 
In Fig.2 we compared eq.(\ref{power law decay}) and an exact numerical
computation of the contour integral along C$_{1}$, 
which shows that the formula (\ref{power law decay}) 
is an excellent approximation.

As an application we used this formula to estimate the transition time
to the power law decay for the $\pi \rightarrow \mu \bar{\nu }_{\mu }$ decay.
We use
\begin{eqnarray}
&&
\sigma (\omega ) = \frac{\Gamma _{\pi }}{2\pi }\,(\frac{\omega }{m_{\pi }})^{2}
\,\frac{(1 - \frac{m_{\mu }^{2}}{\omega ^{2}})^{2}}
{(1 - \frac{m_{\mu }^{2}}{m_{\pi }^{2}})^{2}} \,, 
\\ &&
\Gamma _{\pi } = \frac{1}{8\pi }G_{F}^{2}\,f_{\pi }^{2}m_{\pi }m_{\mu }^{2}\,
(1 - \frac{m_{\mu }^{2}}{m_{\pi }^{2}})^{2} \,.
\end{eqnarray}
The result is
\begin{equation}
t_{p} \approx \frac{2}{\Gamma _{\pi }}\,\ln \left( \,
\frac{\pi }{4}\,(\frac{m_{\pi }}{\Gamma _{\pi }})^{4}\,
(1 - \frac{m_{\mu }}{m_{\pi }})^{2}\,
(1 - \frac{m_{\mu }^{2}}{m_{\pi }^{2}})^{2}\,\right) \approx 
\frac{280}{\Gamma _{\pi }} \,.
\end{equation}
Since $e^{-\Gamma _{\pi }t_{p}}$ is too small, it is difficult to
observe the power law decay in this example.

On the other hand, at early times the high frequency behavior of the
spectral function $\sigma (\omega )$ becomes important, on which
we shall mainly focuss.
One naively expects that the short-time behavior exhibits a deviation
from the exponential law in the form of
\begin{equation}
|\langle 1|e^{-\,iHt}|1 \rangle|^{2} \approx 1 - t^{2}\,
\left( \, \langle 1|H^{2}|1 \rangle - \langle 1|H|1 \rangle^{2} \,\right)
\,.
\end{equation}
Thus, quantum mechanics appears to predict the quadratic form of deviation
in the $t\rightarrow 0$ limit.
More precisely, our solution (\ref{sol for non-decay}) gives
\begin{equation}
|c(t)|^{2} \approx 1 - t^{2}\,\overline{\Delta (\omega - E_{1})^{2}}
\,, \hspace{0.5cm} 
\overline{\Delta A^{2}} \equiv \int_{\omega _{c}}^{\infty }\,d\omega \,
H(\omega )\,A^{2} - \left( \,\int_{\omega _{c}}^{\infty }\,d\omega \,
H(\omega )\,A\,\right)^{2} \,.
\end{equation}
As will be made more explicit shortly, field theory models, worked in
perturbation, give at best
(in the super-renormalizable case) the high energy behavior of
\( \:
H(\omega ) \rightarrow {\rm const}\;/\omega ^{2} \,, 
\: \)
as $\omega \rightarrow \infty $.
The coefficient of order $t^{2}$ term is then infinite in field 
theory.\cite{maiani et al 93} 
It is however not clear whether the power series expansion inside the
$\omega $ integral for estimating $t \rightarrow 0$ limit
is mathematically permissible or not.

\vspace{0.5cm} 
3. {\bf Field theory models and renormalization} 

The projector formalism thus presented recasts the non-decay
amplitude into the form known in an exactly solvable field theory
model, the Lee model.\cite{lee model}
We briefly summarize this connection below.
For general renormalizable field theory models there is however 
a great difference from the Lee model, in that 
the spectral weight $\sigma (\omega )$ in the projector formalism cannot be
arbitrarily given and is fixed by the theory, which however is 
difficult to extract beyond perturbation.
We shall nevertheless use the form of renormalization suggested from
the study of the Lee model 
for unstable particle decay.\cite{decay in lee model} 
Renormalization in a more conventional approach that assumes existence
of the asymptotic field of unstable particle is dealt with in
Appendix.

Let us first note that the projector formalism
may be recast into an equivalent quantum mechanical model.
We imagine that the decay process is a transition between the states,
\( \:
|1\rangle \rightarrow |\omega \rangle \,.
\: \)
All other states that decouple from the decay process may be ignored,
which should not pose any problem for our purpose.
With $E_{1} > \omega _{c} \,,$ 
the Hamiltonian $h$ of this model is of the form, 
\begin{eqnarray}
&& \hspace*{-1cm}
h = 
E_{1} \,|1 \rangle \langle 1| + 
\int_{\omega _{c}}^{\infty }\,d\omega 
\,\omega |\omega \rangle \langle \omega |
+  \int_{\omega _{c}}^{\infty }\,d\omega \,\sqrt{\sigma (\omega )}\,
(\,|\omega \rangle \langle 1| + |1\rangle \langle \omega |\,) 
\,.
\end{eqnarray}
We shall show the equivalence after we introduce a field theory
model.

We assume that the states relevant to the decay process are given by
a Fock space of
\begin{equation}
|1\rangle = a^{\dag }|0\rangle 
\,, \hspace{0.5cm} 
|\omega \rangle = b^{\dag }(\omega )|0\rangle \,.
\end{equation}
We may then extend the above model to 
an equivalent quantum field theory model (EQFM).  
The model is defined by
\begin{eqnarray}
&&
\hspace*{-1cm}
H = (E_{1} + \delta E)\,a^{\dag }a
+ \int_{\omega _{c}}^{\infty }\,d\omega \,
\omega \,b^{\dag }(\omega )b(\omega ) + \frac{1}{\sqrt{Z}}\,
\int_{\omega _{c}}^{\infty }\,d\omega \,\sqrt{\sigma (\omega )}\,
\left( \,ab^{\dag }(\omega ) + a^{\dag }b(\omega )\,\right) \,.
\nonumber \\ &&
\end{eqnarray}
Here $\delta E$ and $Z$ are renormalization constants 
needed to compensate against a bad high frequency behavior
of $\sigma (\omega )$.

This model has an exact conservation law;
the sum of the total $a-$ and $b-$particle numbers is conserved.
This enables one to solve this model in each sector of a given
conserved particle number, separately.
The consequence of the projector formalism follows by identifying
the projector
\( \:
P = a^{\dag }|0\rangle \langle 0|a 
\: \)
in this field theory model.

The freedom of choosing renormalization constants, $Z \,, \delta E$,
are fixed later by a renormalization prescription. 
Allowing for their existence is crucial to renormalizability.
If $Z$ is complex, one may redefine the phase associated with the state
$|1\rangle $ such that $Z$ may be taken to be real and positive.
Although written in a different form, it should be evident that
this model is in essence equivalent to the Lee model for the case of
the vanishing $N$ particle mass, $m_{N} = 0$.
In the Lee model for unstable particle decay this factor
$Z$ is infinite in the limit of local field theory
(when the so called cutoff function is unity).\cite{decay in lee model}

We assume the canonical commutation relation;
\begin{equation}
[\,b(\omega ) \,, b^{\dag }(\omega ')\,] = \delta (\omega - \omega ')
\,, \hspace{0.5cm} 
[\,a\,, a^{\dag } \,] =  1 \,.
\end{equation}
Since the continuous states $|\omega \rangle $ are two-body states
of the decay product, assumption of the canonical commutator for
$b(\omega )$ is highly non-trivial. 
It is conceivable that there is some correction to
this relation. Neglect of such possible correction should be regarded
as an essential part of the approximation to the full theory.
Furthermore, there are parts of the full-fledged Hamiltonian 
such as $a^{2}\,b^{\dag }(\omega )$ that
are not included here, because they are not relevant to one-particle
decay process.
When one discusses many-body problem of unstable particles, they
may become relevant.

We also note that in our EQFM approach we do not assume existence
of asymptotic fields. This is welcome because the asymptotic field
corresponding to unstable particle is not well defined in a strict
sense.

To show that time evolution of the state $|1\rangle $ in EQFM is
equivalent to the one in the projector formalism, 
we first seek energy eigenstate $|\omega \rangle _{S}$ with
\( \:
(H - \omega )\,|\omega \rangle _{S} = 0 \,,
\: \)
in the form of scattering state; we find that
\begin{eqnarray}
&& \hspace*{0.5cm} 
|\omega \rangle _{S} = B^{\dag }(\omega )|0\rangle \,, 
\\ && \hspace*{-2cm}
B^{\dag }(\omega ) = b^{\dag }(\omega ) +
F(\omega + i0^{+})\,\left( \,-\,\sqrt{\sigma (\omega )}\,
\frac{a^{\dag }}{\sqrt{Z}}
+ \frac{1}{Z}\,\int_{\omega _{c}}^{\infty }\,d\omega' \,
\frac{\sqrt{\sigma (\omega )\sigma (\omega ')}}{\omega ' - \omega - i0^{+}}
\,b^{\dag }(\omega ')\,\right) \,, 
\\ &&
F(z)^{-1} = -\,z + E_{1} + \delta E
- \frac{1}{Z}\,\int_{\omega _{c}}^{\infty }\,d\omega \,
\frac{\sigma (\omega )}{\omega - z} 
\,.
\end{eqnarray}
There exists an important relation of Hamiltonian equivalence,
\begin{equation}
H = \int_{\omega _{c}}^{\infty }\,d\omega \,\omega \,B^{\dag }(\omega )
B(\omega ) \,.
\end{equation}
Furthermore, two descriptions, either in terms of $(a \,, b(\omega ))$ or
$(B(\omega ))$, are related by a canonical transformation,
satisfying 
\begin{eqnarray}
&&
[\,B(\omega ) \,, B^{\dag }(\omega ')\,] = \delta (\omega - \omega ')\,.
\end{eqnarray}

We further note a remarkable operator inversion;
\begin{eqnarray}
&&
a^{\dag } = -\,\frac{1}{\sqrt{Z}}\,\int_{\omega _{c}}^{\infty }\,d\omega \,
\sqrt{\sigma (\omega )}\,F^{*}(\omega + i0^{+})\,B^{\dag }(\omega ) 
\,, 
\\ &&
b^{\dag }(\omega ) = B^{\dag }(\omega ) + \frac{1}{Z}\,
\int_{\omega _{c}}^{\infty }\,d\omega' \,
\frac{\sqrt{\sigma (\omega )\sigma (\omega ')}\,F^{*}(\omega ' + i0^{+})}
{\omega - \omega ' + i0^{+}}\,B^{\dag }(\omega ') \,.
\end{eqnarray}

The proof of all these is based on the analytic property of $F(z)$ having
the cut on the real axis from $z > \omega _{c} $, 
with the discontinuity formula of
\begin{equation}
F(\omega + i0^{+}) - F(\omega - i0^{+}) = \frac{2\pi i}{Z}\sigma (\omega )\,
|F(\omega + i0^{+})|^{2}
\,, \hspace{0.5cm} {\rm for} \;\omega > \omega _{c} \,.
\end{equation}
Its asymptotic behavior,
\( \:
F(z) \rightarrow -\,\frac{1}{z} \,,
\: \)
is also important.
This asymptotic behavior holds only when a finiteness condition for
\begin{eqnarray}
&&
E_{1} + \delta E - \frac{1}{Z}\,\int_{\omega _{c}}^{\infty }\,d\omega \,
\frac{\sigma (\omega )}{\omega } \,, 
\\ &&
\frac{\omega }{Z}\,\int_{\omega _{c}}^{\infty }\,d\omega '\,
\frac{\sigma (\omega ')}{\omega '(\omega ' - \omega - i0^{+})} \,, 
\end{eqnarray}
is fulfilled.

Time evolution of the state $|1\rangle $ is derived, by noting the
completeness \cite{glaser-kallen};
\begin{eqnarray}
&&
\int_{\omega _{c}}^{\infty }\,d\omega \,
|\omega \rangle _{S}\,_{S}\langle \omega | =
|1 \rangle \langle  1| + 
\int_{\omega _{c}}^{\infty }\,d\omega \,
|\omega \rangle \langle \omega | \,.
\end{eqnarray}
We find from
\begin{equation}
|1 \rangle = \int_{\omega _{c}}^{\infty }\,d\omega \,_{S}\langle \omega |
1 \rangle\,|\omega \rangle _{S} \,, 
\end{equation}
that
\begin{eqnarray}
&&
e^{-\,iHt}\,|1 \rangle =
-\,\frac{1}{\sqrt{Z}}\,
\int_{\omega _{c}}^{\infty }\,d\omega \,\sqrt{\sigma (\omega )}\,
F^{*}(\omega + i0^{+})\,e^{-i\omega t}\,|\omega \rangle_{S} \,, 
\\ &&
\langle 1|e^{-iHt}|1 \rangle = \langle 0|a\,e^{-iHt}\,a^{\dag }|0 \rangle
= \frac{1}{Z}\,\int_{\omega _{c}}^{\infty }\,d\omega\,\sigma (\omega )\,
|F(\omega + i0^{+})|^{2}\,e^{-i\omega t} \,.
\end{eqnarray}
Due to the cut structure of $F(z)$, one can write this as
\begin{eqnarray}
\langle 1|\,e^{-\,iHt}\,|1 \rangle &=& 
\frac{1}{2\pi i}\,\int_{-\infty }^{\infty }\,d\omega \,
e^{-i\omega t}\,F(\omega + i0^{+}) 
\\ 
&=& \frac{1}{Z}\,
\int_{\omega _{c}}^{\infty }\,d\omega \,e^{-i\omega t}\,
\frac{\sigma (\omega )}{|D(\omega )|^{2}} 
\,, \label{ren non-decay amp} 
\\ 
|D(\omega )|^{2} &=& \left( \,
\omega - E_{1} - \delta E + \frac{1}{Z}\,
{\cal P}\,\int_{\omega _{c}}^{\infty }\,d\omega '\,\frac{\sigma (\omega ')}
{\omega ' - \omega }
\,\right)^{2}
+ \frac{1}{Z^{2}}(\,\pi \sigma (\omega )\,)^{2} \,.
\end{eqnarray}
This result coincides with the result of the projector formalism
(\ref{sol for non-decay}) except for the $Z$ factor and $\delta E$.

There are various ways of how to specify the renormalization condition.
We adopt an interesting proposal using 
the residue of the physical pole of unstable particle 
in ref \cite{decay in lee model};
\( \:
|F(z)| \sim  N |\frac{1}{z - E_{*}}| \,, 
\: \)
where $E_{*} = E_{r} - iE_{i}$ is the pole location in the second sheet,
and $N$ is some finite constant.
The renormalization condition then reads as
\begin{eqnarray}
&&
\hspace*{-2cm}
\left( \,1 + \frac{1}{Z}\,
\int_{\omega _{c}}^{\infty }\,d\omega \,
\sigma (\omega )\,\frac{(\omega - E_{r})^{2} - E_{i}^{2}}
{|\omega - E_{*}|^{4}}\,\right)^{2} 
+ \frac{4}{Z^{2}}\,\left( \,
\int_{\omega _{c}}^{\infty }\,d\omega \,
\sigma (\omega )\,\frac{E_{i}(\omega - E_{r})}{|\omega - E_{*}|^{4}} 
\,\right)^{2} = N^{-2} \,. \label{ren condition} 
\end{eqnarray}
For a given high frequency behavior of $\sigma (\omega )$ 
renormalizability demands that all infinities, if there is any,
must be eliminated
by two infinite constants, $\delta E$ and $Z$, 
considering that the renormalized and observable quantity $E_{*}$ is finite.
The quantity $Z\,|D(\omega )|^{2}$ appearing
in the non-decay amplitude (\ref{ren non-decay amp})
should then be finite in renormalizable theories,
possibly with an exception of an infinite phase factor 
that does not contribute to the non-decay probability.
The renormalization condition (\ref{ren condition}) may be called
the on-shell renormalization condition.

When the integral 
\begin{equation}
A_{1} \equiv 
\int_{\omega _{c}}^{\infty }\,d\omega \,\frac{\sigma (\omega )}{\omega ^{2}}
\end{equation}
is finite, the infinity appears only in the mass shift, $\delta E$.
On the other hand, when this integral is infinite, one finds that
\begin{eqnarray}
&&
Z \approx \frac{N}{1 - N}\,
\int_{\omega _{c}}^{\infty }\,d\omega \,
\sigma (\omega )\,\frac{(\omega - E_{r})^{2} - E_{i}^{2}}
{|\omega - E_{*}|^{4}} \,. 
\end{eqnarray}
The criterion for existence of the remaining infinity $Z$
is that the integral $A_{1}$ is infinite.
We note that the on-shell renormalization
condition (\ref{ren condition}) is necessarily
non-perturbative, in the sense that the solution $Z$ does not behave like
\( \:
Z = 1 + O[\sigma ] \times 
\: \)
an infinity, as might have been expected.
If $N \approx 1 + O[\sigma ]$ \,, 
it behaves rather like
\( \:
Z = O[\sigma^{0} ] \times 
\: \)
an infinity.

We now show some examples of the spectral function $\sigma (\omega )$ in
perturbation.
Three models are taken;
a superrenormalizable model of boson decay defined by an interaction
Lagrangian density,
\( \:
{\cal L}_{b} = \frac{\mu }{2}\,\varphi \chi ^{2} \,, 
\: \)
a renormalizable model of fermion-pair decay,
\( \:
{\cal L}_{f} = g\,\varphi \bar{\psi }\psi \,,
\: \)
and an effective model of proton decay,
\( \:
{\cal L}_{p} = \bar{e}(a + b\gamma _{5})p\,\pi + ({\rm h.c.})\,, 
\: \)
where $a \,, b$ are constants of order,
\( \:
\frac{g_{X}^{2}m_{p}^{2}}{m_{X}^{2}} \,.
\: \)
Here $m_{X}$ is the mass ($\approx 10^{16}\,$ GeV)
of heavy $X$ boson mediating baryon non-conservation.
Let us first note that in the projector formalism, or more simply in EQFM, 
there is no Z-diagram as shown in Fig.3, contributing, and one
has to consider the direct diagram alone. 
Perturbative computation in the coupling, $\mu $  and $g$, 
then gives the spectral function $\sigma (\omega )$ to lowest order;
\begin{eqnarray}
&&
\sigma _{b}(\omega ) =  \frac{\mu ^{2}}{64\pi ^{2}\,M}\,
\sqrt{1 - \frac{4m^{2}}{\omega ^{2}}} \,, 
\\ &&
\sigma _{f}(\omega ) = 
\frac{g^{2}}{16\pi ^{2}\,M}\,\omega ^{2}
\left(1 - \frac{4m^{2}}{\omega ^{2}}\right)^{3/2} \,,
\\ &&
\sigma _{p}(\omega ) =
\frac{|a|^{2} + |b|^{2}}{32\pi ^{2}}\,\omega \,
\left( 1 - \frac{m^{2}}{{\omega }^{2}}\right)^{2} \,.
\end{eqnarray}
Here $M$ and $m$ denote the parent mass and the daughter mass,
respectively.
(In the case of the proton decay we can take the electron mass vanishing
such that $m$ is the pion mass.)
The lowest threshold $\omega _{c} = 2m$ (or $= m$ for the proton decay).

We point out that in the case of fermion-pair decay there is
a quadratic divergence for the self-energy diagram of
\( \:
\varphi \rightarrow \psi \bar{\psi } \rightarrow \varphi \,, 
\: \)
hence the spectral function increases as 
\( \:
\sigma _{f}(\omega ) \propto \omega ^{2}
\: \)
for
\( \:
\omega \rightarrow \infty \,.
\: \)
This bad high energy behavior in the renormalizable model introduces 
infinities in the integral,
\( \:
\int_{\omega _{c}}^{\infty }\,d\omega \,
\frac{\sigma (\omega )}{\omega  - z } \,
\: \)
for $G(z)$, proportional to
\( \:
z^{0} \,,\; z \,,\; z ^{2} \,, 
\: \)
and the term independent of $z$ is eliminated by the mass renormalization.
The remaining integral contains an infinite $A_{1}$ term, and
the renormalizability of this particular field theory model is determined
whether
\begin{eqnarray}
&&
\frac{1}{A_{1}}\,{\cal P}\,\int_{\omega _{c}}^{\infty }\,d\omega '\,
\frac{\sigma (\omega ')}{\omega '(\omega ' - \omega )}
\end{eqnarray}
is finite or not.
The fermion-pair decay model here indeed obeys this finiteness condition.

In Appendix we show that if one takes the initial one-particle state
at rest given by
\begin{equation}
|1\rangle = -\,i\,\int\,d^{3}x\,f_{0}
(\vec{x} \,, t_{i})\stackrel{\leftrightarrow }
{\partial }_{0}\varphi (\vec{x} \,, t_{i})\,|0\rangle \,, 
\hspace{0.5cm} 
f_{0}(\vec{x} \,, t_{i}) = \frac{e^{-iMt_{i}}}{\sqrt{2M}} \,, 
\end{equation}
where $\varphi (x)$ is the Heisenberg operator of unstable particle
(whose existence is assumed) and $|0\rangle $ is the true vacuum state,
then the non-decay amplitude is related to the ordinary Green's function;
\begin{eqnarray}
&& \hspace*{-1cm}
\langle 1|e^{-\,iH(t - t_{i})}|1 \rangle = 
\frac{(i\partial _{t} + M)^{2}}{2M}\,
\int\,d^{3}x\,\int\,d^{3}x_{i}\,
\langle 0|T(\varphi (x)\varphi (x_{i}))|0 \rangle
\,. 
\end{eqnarray}
Assuming a convergence of integral, one can write this as
\begin{eqnarray}
-\,\frac{1}{2\pi i}\,\int_{-\infty }^{\infty }\,
d\omega \,\frac{(\omega + M)^{2}}{2M}\,
e^{-i\omega (t - t_{i})}\, \Delta (\omega ^{2} + i0^{+}) 
\,,
\end{eqnarray}
where $\Delta $ is the usual Green's function expressed in terms of
the Lehmann spectral function $\rho _{L}(\sigma ^{2})$;
\begin{equation}
\Delta (\omega ^{2}) = -i\,\int\,d^{4}x\,
e^{i\omega t}\,\langle 0|T(\varphi (x)\varphi (0))|0 \rangle 
=
\int_{\omega _{c}^{2}}^{\infty }\,
d\sigma ^{2}\,\frac{\rho _{L}(\sigma  ^{2})}
{\omega ^{2} - \sigma ^{2} + i0^{+}} 
\,.
\end{equation}
It should be evident that the non-decay amplitude is finite 
after renormalization if the Green's function is renormalizable.
This is in sharp contrast to the argument based on perturbative
expansion \cite{maiani et al 93}.
We shall discuss the problem of renormalization in the field
theoretical framework further in Appendix.
It is also true that the 
disconnected diagram is factorized in the non-decay amplitude,
and one may ignore the disconnected diagram in subsequent computation.

\vspace{0.5cm} 
4. {\bf Field theory with finite time resolution}

To search for deviation from the exponenital decay at short times,
one should keep in mind that following time evolution with infinitely
good time resolution may not be meaningful and certainly is not
practical with a limited accuracy of time measurement.
With this in mind we consider how the decay proceeds with a finite
time resolution.
As mentioned in Introduction, we take the resolution in the range,
\( \:
\frac{1}{\Gamma } > \Delta t > \frac{1}{M} \,.
\: \)
With this coarse graining one needs the spectral function $\sigma (\omega )$
only in a relatively low frequency range, and the high frequency part is
effectively cut off by a finite time resolution.
This makes it possible to discuss the decay law even for
non-renormalizable theories such as the four-Fermi type theory given by
$X$ boson mediated interaction.

We time average the basic function $\beta (t - t')$
using a resolution function $\tilde{\delta }(t \,; \Delta t)$,
\begin{eqnarray}
&& \hspace*{-1cm}
\tilde{\delta }(t \,; \Delta t ) \approx  0 \hspace{0.3cm} {\rm for} \;
|t| \gg \Delta t \,; \hspace{0.5cm} 
\left( \,
\frac{d}{dt}\,\tilde{\delta }(t \,; \Delta t)
\,\right)_{t = 0} = 0 \,, 
\hspace{0.5cm} 
\int_{-\infty }^{\infty }\,dt\,\tilde{\delta }(t\,; \Delta t) = 1 \,.
\nonumber \\ &&
\end{eqnarray}
Examples of the resolution function are the step function and
the Gaussian function,
\begin{eqnarray}
&&
\hspace*{-1cm}
\tilde{\delta }^{{\rm S}}(t \,; \Delta t ) 
= \frac{\theta (\,\frac{\Delta t}{2}- |t|\,)}{\Delta t} \,, \hspace{0.5cm} 
\tilde{\delta }^{{\rm G}}(t \,; \Delta t ) 
= \frac{1}{\sqrt{2\pi }\,\Delta t }\,e^{-\,\frac{t^{2}}{2\,(\Delta t)^{2}}} 
\,.
\end{eqnarray}
Another useful resolution function is the inverse Fourier transform
of a Lorentzian form,
\begin{equation}
\delta _{F}^{L}(\omega \,; \Delta t) = \frac{1}{(1 + \omega ^{2}
\,\Delta t^{2})^{2}} \,, \label{lorentzian resolution} 
\end{equation}
giving in the real time 
\begin{equation}
\tilde{\delta }^{L} (t \,; \Delta t) = \frac{1}{\Delta t}\,
(1 + \frac{2|t|}{\Delta t})\,e^{-2|t|/\Delta t} \,.
\end{equation}
With a finite resolution, the averaged function is
\begin{eqnarray}
\overline{\beta (\tau )} &\equiv& \int_{-\infty }^{\infty }\,d\tau '\,
\beta (\tau ')\tilde{\delta }(\tau ' - \tau \,; \Delta t)
\nonumber 
\\ 
&=& \int_{\omega _{c}}^{\infty }\,d\omega \,
\overline{e^{-i(\omega - E_{1})\tau }} \,\sigma (\omega ) \,.
\end{eqnarray}

By defining the Fourier transform of the resolution function,
\begin{equation}
\delta _{F}(\omega \,; \Delta t) = \int_{-\infty }^{\infty }\,dt\,
e^{-i\omega t}\,\tilde{\delta }(t\,; \Delta t ) \,, 
\end{equation}
one finds that
\begin{eqnarray}
&&
\overline{G(\omega )} = \int_{\omega _{c}}^{\infty }\,d\omega '\,
\sigma (\omega ')\,
\frac{\delta _{F}(\omega ' - E_{1} \,; \Delta t) }{\omega ' - \omega 
- i0^{+}}
\,.
\end{eqnarray}
This formula has a simple interpretation;
with a finite time resolution
the original spectral function $\sigma (\omega )$ is modified to
\( \:
\sigma (\omega )\,\delta _{F}(\omega - E_{1} \,; \Delta t) \,.
\: \)
Thus, the direct diagram has a spectral function peaked at $E_{1}$ of
width $1/(2\Delta t)$, which is cut off at the threshold $\omega _{c}$.

Instead of
\( \:
\beta (t - t') = \langle 1|V_{I}(t)\,V_{I}(t')|1\rangle \,, 
\: \)
one may introduce the finite resolution both at
$V_{I}(t)$ and at $V_{I}(t')$ using the Fourier transformed resolution function
$\delta _{F}(\omega \,; \Delta t)$.
In this case the Fourier transform of the resolution function 
for $\beta (t - t')$ is
\( \:
(\delta _{F}(\omega  \,; \Delta t))^{2} \,;
\: \)
the resolution function is modified, but it is still of the form
of the resolution for $\beta (t - t')$.
Thus one may view our averaging as a coarse graining of local operators
in quantum field theory.

\vspace{0.5cm} 
5. {\bf Application to unstable particle decay} 

A universal effect of the resolution $\delta _{F}$ is to cut off the
high frequency contribution in the region of
\( \:
|\,\omega -  M\,| > \frac{1}{\Delta t} \,.
\: \)
Furthermore, if 
\( \:
\Delta t \geq  \frac{1}{2Q} \,, 
\: \)
the contribution near the threshold $\omega \approx Q$ is also cut off.
The effect of finite resolution then prolongs the epoch of the exponential
decay, since the threshold behavior responsible at late times becomes
less important.
We shall estimate how deviation from the exponential law occurs at early times,
assuming 
\( \:
\frac{1}{M} \leq \Delta t \ll \frac{1}{\Gamma }
\: \)
(decay lifetime).
Obviously, for $\Delta t \geq  \frac{1}{\Gamma }$ the decay law is
complicated and does not obey the simple exponential law; in this case one
has to make indivisual analysis taking into account each specific situation.

We first compute the amplitude of the exponential decay when this law
holds.
The magnitude of the exponential decay is determined by the residue of
the pole in the second Riemann sheet.
For this purpose we may use an approximate formula for the real part
of the self-energy;
\begin{equation} 
\frac{\Gamma }{2\pi }\,{\cal P}\,
\int_{2m}^{\infty }\,dx\,\frac{\delta _{F}(x - M\,; \Delta t)}
{x - \omega } \,,
\end{equation}
where $\Gamma $ is the decay rate in perturbation theory.
We shall take a simplified form of resolution for analytical estimate,
\begin{equation}
\delta _{F}(x\,; \Delta t) = \theta (\,\frac{1}{2\Delta t} - |x|\,) \,.
\end{equation}
For small $g$ the pole term is then approximately given by
\begin{equation}
F(z) \approx -\,\frac{(\,1 - \frac{2}{\pi }\,\Gamma \Delta t\,)^{-1}}{z -
M + i\,\frac{\Gamma }{2}} \,.
\end{equation}
We have ignored a small imaginary part of the residue, which is of
order $\Gamma /M$.
The residue factor 
\( \:
R = (\,1 - \frac{2}{\pi }\,\Gamma \Delta t\,)^{-1}
\: \)
is replaced by
\( \:
R = (\,1 - \frac{\Gamma }{2\pi \,Q}\,)^{-1}
\: \)
for $\Delta t < \frac{1}{2Q}$.
The exponential period is approximately described by the probability
function,
\begin{equation}
|c(t)|^{2} \approx R^{2}\,e^{-\,\Gamma t} \,, \hspace{0.5cm} 
R^{-1} = 1 - \frac{2\Gamma }{\pi }\,
{\rm Max}\;(\,\Delta t \,, \frac{1}{4Q} + \frac{\Delta t}{2}\,)
\,.
\end{equation}
The enhanced magnitude $1 - R^{2}$ becomes maximal around
\( \:
\Delta t = \frac{1}{2Q} \,, 
\: \)
with a value around $\frac{\Gamma }{\pi Q}$.
Although time variation of
the enhanced exponential decay is difficult to distinguish from the
ordinary exponential law,
it may be possible to observe the magnitude of the enhanced non-decay
probability if one can measure the onset time of decay.

Let us estimate the onset time of the exponential period, first by
examining the non-decay probability at early times.
Short time behavior of the non-decay probability is most sensitive to
the high frequency behavior of $\sigma (\omega )$.
By expanding $e^{-i(\omega - M)t}$, one obtains for a small $t$
\begin{eqnarray}
\hspace*{-1cm}
|c(t)|^{2} &\approx& 1 - t^{2}\,
\left[ \,\int_{2m}^{\infty }\,d\omega \,H_{R}(\omega )\,(\omega - M)^{2}
- \left( \int_{2m}^{\infty }\,d\omega \,H_{R}(\omega )
\,(\omega - M)\right)^{2}\,\right] \,, \label{short-time formula} 
\\ &&
H_{R}(\omega ) = 
\frac{\sigma (\omega )\,\delta _{F}(\omega \,; \Delta t)}
{(\omega - E_{r} + \Pi_{R} (\omega ))^{2}
+ (\pi \sigma (\omega)\delta _{F}(\omega \,; \Delta t))^{2}} \,, 
\end{eqnarray}
where $\Pi _{R}$ is determined by the dispersion relation from
$\sigma (\omega)\delta _{F}(\omega \,; \Delta t)$.
As a crude estimate one may take a truncated Breit-Wigner form,
\begin{equation}
H_{R}(\omega )
= \frac{\Gamma }{2\pi }\,\frac{\theta (\,\frac{1}{2\Delta t} -
|\omega - M|\,)}{(\omega - M)^{2} + \frac{\Gamma ^{2}}{4}} \,,
\label{truncated bw} 
\end{equation}
which can be used only when the $Q$ value ($= M - 2m$) is not
too small;\\
\( \:
2(M - 2m)\,\Delta t > 1 \,.
\: \)
With $\Gamma \Delta t \ll 1$,
\begin{equation}
|c(t)|^{2} \approx 1 - \frac{\Gamma }{2\pi \,\Delta t}\,t^{2} \,.
\end{equation}
Thus, the coefficient of order $t^{2}$ term formally diverges linearly as
\( \:
\Delta t \rightarrow 0 \,.
\: \)
Since the power series expansion in time $t$ is valid only for
$t < \Delta t$, this does no harm practically.
This simplified short-time formula is reasonably good, but
the approxiamtion is not excellent, presumably reflecting dependence
on a precise form of resolution function taken.

More quantitative questions can be raised as to how the decay law is 
changed if one varies the resolution within physics mass scale,
\( \:
\Delta t \geq  \frac{1}{M} \,
\: \)
($M$ being the parent mass).
We have numerically computed the non-decay probability for the boson-pair
and fermion-pair decay models, assuming various time resolution $\Delta t$.
Examples of the result of this numerical computation 
for the case of the fermion-pair decay are presented in Fig.4 and Fig.5.
As qualitatively discussed above, deviation from the exponential law appears
at short times of order $\Delta t$ or less. At the same time the exponential
term in intermediate times is enhanced by $\Delta t$ dependent amount.
In Fig.4 we show the non-decay probability in the entire time range
including three different time stages;
early $1 - O[t^{2}]$ in the inlet of the figure, 
intermediate exponential, and late power stages.
In Fig.5 we show the short-time behavior in comparison to the limiting
behavior; $1 - O[t^{2}]$ region using eq.(\ref{short-time formula}),
and the exponential region using a numerically computed pole location
and its residue.
In all these numerical computations we used the Lorentzian form of
time resolution, eq.(\ref{lorentzian resolution}).

The transition time $t_{*}$ to the exponential period is crudely estimated
by equating this short-time formula to the enhanced exponential formula;
\( \:
1 - \frac{\Gamma t_{*}^{2}}{2\pi \Delta t} = R^{2}(1 - \Gamma t_{*})\,.
\: \)
It gives, with
\( \:
R^{2} \approx 1 + \frac{4}{\pi }\,\Gamma \Delta t \,, 
\: \)
\begin{equation}
t_{*} \approx (\pi - \sqrt{\pi ^{2} - 8})\,\Delta t \sim 1.8\,\Delta t
\,.
\end{equation}
In our numerical computation this crossover of the two formulas of
the regions does not
actually occur. Nevertheless, the transition to the exponential period
is given by a factor of a few times $\Delta t$, as seen from Fig.5.

In estimating the short time limit of the non-decay amplitude,
one needed the high energy behavior of the spectral function 
$\sigma (\omega )$.
Let us consider a cutoff theory of a finite time resolution $\Delta t$
in which the $A_{1}$ integral diverges in the $\Delta t \rightarrow 0$
limit.
If the time resolution $\Delta t$ is small but is still much larger than
the inverse of the cutoff scale $1/\Lambda $ of physical relevance, 
one may discuss the short-time behavior 
in the range $t > \frac{1}{\Lambda }$.
This consideration may be applied to non-renormalizable models of
four-Fermi type, where $\Lambda $ is the weak or the $X$ boson mass.
The use of perturbative formula for the spectral function $\sigma (\omega )$
limits the allowed range of the resolution $\Delta t$, since the integral
$A_{1}$ must remain of order unity.
We shall use this constraint to examine the short-time limit of
$t \approx \Delta t$, at which one may expect a deviation from
the exponential law.

For definiteness, we considered four examples; the $\mu $ decay,
\( \:
\mu \rightarrow \nu _{\mu }\,e\bar{\nu }_{e} \,, 
\: \) 
the neutron decay,
\( \:
n \rightarrow p\,e\bar{\nu }_{e} \,, 
\: \)
the $\pi $ decay,
\( \:
\pi \rightarrow l\bar{\nu }_{l} \,,
\: \)
and the proton decay,
\( \:
p \rightarrow e^{+}\pi ^{0} \,.
\: \)
In all these cases we find, assuming $\Delta t> \frac{1}{2Q} $ and
$\gg \frac{1}{2M}$, a universal result;
\begin{equation}
A_{1} \approx \frac{\Gamma_{i} }{2\pi M_{i}^{2}\,\Delta t} \,, 
\end{equation}
since 
\begin{equation}
\frac{\sigma _{i}(\omega )}{\omega ^{2}} = \frac{1}{2\pi M_{i}}\,
\frac{\Gamma _{i}(\omega )}{\omega }
\end{equation}
may be approximated by 
\( \:
\frac{\Gamma_{i}}{2\pi M_{i}^{2}}
\: \)
in the $A_{1}$ integrand.
Hence, the condition
\( \:
A_{1} < 1 \,, 
\: \)
gives
\begin{equation}
\Delta t > \frac{1}{2\pi }\,\frac{\Gamma_{i} }{M_{i}^{2}} 
= \frac{1}{\Gamma_{i} }\,\frac{1}{2\pi }\,(\frac{\Gamma_{i} }{M_{i}} )^{2}
\,. 
\end{equation}
Thus, the exponential law is good as early as the lifetime times
of order,
\( \:
(\frac{\Gamma_{i} }{M_{i}} )^{2} \,,
\: \)
which is very small in all cases.

Let us examine more closely the case of a small $Q$ value;
the neutron decay for $\Delta t < \frac{1}{2Q}$ and
a hypothetical case of $\pi \rightarrow 
l\bar{\nu }_{l}$ decay, in which the mass difference is very small;
\( \:
m_{\pi } - m_{l} \ll m_{\pi } \,.
\: \)
These are special examples of the small $Q$ limit.
First, let us discuss the $\pi $ decay.
With $Q = m_{\pi } - m_{l}$ and for the case of
\( \:
\Delta t < \frac{1}{2Q} 
\: \), 
one can work out the constraint $A_{1} < O[1]$, to get
\begin{equation}
\frac{G_{F}^{2}\,f_{\pi }^{2}m_{l}^{2}}{16\pi ^{2}\,m_{\pi }}\,
\int_{m_{l}}^{m_{\pi } + 1/(2\Delta t)}\,d\omega \,
(1 - \frac{m_{l}^{2}}{\omega ^{2}})^{2}
< O[1] \,.
\end{equation}
Because a typical dimensionless quantity in the left hand side is
$\approx G_{F}^{2}\,m_{\pi }^{4} \approx 10^{-14}$, this constraint
is readily obeyed for $\Delta t > 1/m_{\pi }$.

In the case of the neutron decay for a small $\Delta t$, one has,
assuming $\Delta t < \frac{1}{2Q}$,
\begin{eqnarray}
&& \hspace*{-2cm}
\frac{G_{F}^{2}}{2\pi ^{4}}\,
\int_{m_{e}}^{m_{n} - m_{p} + 1/2\Delta t}
\,dE\,E\sqrt{E^{2} - m_{e}^{2}}\,
\int_{E + m_{p}}^{m_{n} + 1/2\Delta t}\,d\omega \,\frac{(\omega 
- E - m_{p})^{2}}{\omega^{2}} 
< O[1] \,.
\end{eqnarray}
It should be clear that even for $\Delta t$ as small as $1/m_{n}$
the integral is of order,
\( \:
m_{n}^{4} \,, 
\: \)
(we are extrapolating a non-relativistic formula to the relativistic
region, which however should be allowed for an order of magnitude
estimate),
hence the quantity in the left hand side is at most of order,
\( \:
G_{F}^{2}\,m_{n}^{4} \approx 10^{-10} \,, 
\: \)
which means that the inequality is violated only at $\Delta t$ unrealistically
small. 
It is thus difficult to observe deviation from the exponential decay
in the short-time region of the neutron decay.

The difficulty of observing deviation from the exponential decay
in realistic particle decay lies in a slow rise of the spectral function
$\sigma (\omega )$ near the threshold;
\( \:
\sigma (\omega ) \propto (\omega - \omega _{c})^{\alpha }
\: \)
with $\alpha \geq 2$.

An implication of this consideration to the nucleon decay is
that underground experiments performed so far can provide meaningful limits
by using the limiting form 
\( \:
1 - \Gamma t
\: \)
of the exponential law.
This holds if the resolution time $\Delta t > 1/m_{N}$.
A caveat for this conclusion is environment effect of hadronic matter,
which will be discussed elsewhere.

Finally, we mention that the infinite time resolution limit,
$\Delta t \rightarrow 0$, can be dealt with in a conventional
approach of relativistic field theory assuming existence of 
the asymptotic field.
The short-time limit $Mt \rightarrow 0$ of the formalism in Appendix however
gives a peculiar behavior of non-decay amplitude
if perturbative formula for the proper
self-energy is used with ordinary renormalization.
It is not clear how non-perturbative effect modifies this result.
On the other hand, both the exponential and the late power behavior
in the approach of Appendix is practically identical to
the one in this section, if the theory is renormalizable. 
A great advantage of introducing
a finite time resolution here is that one can discuss the non-renormalizable
interaction on the same footing as the renormalizable one.

A physical reason why the quadratic decrease in the short-time limit
is obtained with a coarse time resolution
is that a finite
time resolution effectively cuts off the high frequency component, thus
its outcome is similar to the behavior of a system of 
finite number of degrees of freedom.
A field theory of infinitely many degrees of freedom may exhibit
a behavior quite different from quantum mechanics of a finite system.
Unfortunately, the true limit of $t \ll e^{-\,4\pi^{2} /g^{2}}/M$
is difficult to reach in perturbation theory.
One may only probe the region of $t \gg e^{-\,4\pi^{2} /g^{2}}/M$ in
perturbation theory.

\vspace{0.5cm} 
6. {\bf S-wave decay}

We have discussed so far the case of $Q $ not too small.
The small $Q$ limit may be of some interest in relation to the S-wave
decay. 
The decay rate $\Gamma $ goes with small $Q$ as
\( \:
Q^{l + 1/2} \,, 
\: \)
where $l$ is the angular momentum involved in the two particle decay.
Thus the S-wave boson-pair decay (superrenormalizable model here)
gives $\Gamma /Q \propto Q^{-1/2}$ for a small $Q$ and may become
large.
More precisely, the boson-pair decay in the $Q \rightarrow 0$ limit gives
\begin{equation}
\Gamma _{b} \approx \frac{\sqrt{2}g^{2}}{32\pi }\,\sqrt{MQ} \,.
\end{equation}
On the other hand, the P-wave fermion-pair decay
(renormalizable model) gives
\( \:
\Gamma /Q \propto Q^{1/2} \,.
\: \)
Thus, the pole locations move with the Q-value, 
quite differently in the two cases
of the S-wave and the P-wave decay, as shown in Fig.6.
For the exponential decay to exist, the imaginary part of the pole
$\approx \Gamma /2 \ll Q$, and this condition 
may be violated for the S-wave decay of a very small $Q < \Gamma $ value.

In Fig.7 we show an example of the behavior of time evolution
for the S-wave boson-pair decay.
The time evolution shown here corresponds to the parameter set
marked in Fig.6 by the crossed box.
With a very small $Q$, the exponential form is not supported in any
period.
Indeed, the intermediate-time behavior in this case joins  smoothly to
a refined late-time formula of the power behavior.
A refined calculation of the magnitude of the power law is necessary, 
because the $Q\rightarrow 0$ limit gives a constant term in the formula of
$F(\omega )^{-1}$ and the $Q$ in eq.(\ref{power law decay}) is
replaced by another linear term, 
although this constant is small with a small coupling.
The late time formula is realized in this example 
at a time, a few times $1/\Gamma $,
where $\Gamma $ is given either by a naive perturbative formula or by
a half of the imaginary part of the pole (both are roughly of the same
order).
It is however difficult to give a closed form formula of the transition time
to the late power behavior, since the amplitude is not readily
expressible in an analytic form.
Nevertheless, the power of time in the late time formula is unambiguously
determined by the threshold rise of the spectrum,
\( \:
\sigma (\omega ) \propto (\omega - \omega _{c})^{\alpha } \,.
\: \)
It is an interesting question whether this type of behavior may be
found in nature.

\vspace{0.5cm} 
In summary, we have formulated a non-perturbative definition of 
renormalization for the non-decay amplitude of unstable particle decay.
We then discussed how the decay proceeds when viewed with finite
time resolution.
The exponential law usually holds as early as at $\approx $ the resolution
time. 
Exceptions may occur when the S-wave decay rate $> Q$, for which
the power law may appear without the intermediate stage of the
exponential decay.

\vspace{1cm}
\begin{center}
{\bf Acknowledgment}
\end{center}

This work has been supported in part by the Grand-in-Aid for Science
Research from the Ministry of Education, Science and Culture of Japan,
No. 08640341.
The work of the author (I.J.) is supported by a fellowship of
the Japan Society of the Promotion of Science.

%\vspace{0.5cm} 
\newpage
\begin{center}
{\bf Appendix. Non-decay amplitude and Green's function}
\end{center}

\vspace{0.5cm} 
We first argue on general grounds that the non-decay amplitude
should be finite in renormalizable theories.
For this purpose let us recall the Lehmann type of spectral representation
valid in any models.
Using a complete set of eigenstates of the total $H$, one may write
for an arbitrary initial state $|1\rangle $,
\begin{eqnarray}
&&
\langle 1|e^{-\,iHt}|1 \rangle 
= \int_{\omega _{c}}^{\infty }\,d\omega \,\rho (\omega )e^{-\,i\omega t}
 \,, \\
&&
\rho (\omega ) = 
\sum_{n}\,\delta (\omega - E_{n})\,|\langle n|1 \rangle|^{2}
\,.
\end{eqnarray}
Here $|1\rangle $ is not an eigenstate of the total $H$, while 
$|n\rangle $ is. 
Using the analytic function,
\begin{equation}
\tilde{\Delta }  (z) = 
\int_{\omega _{c}}^{\infty }\,d\omega \,\frac{\rho (\omega )}
{\omega  - z} \,, 
\end{equation}
that has the branch cut starting from $z = \omega _{c}$, the non-decay
amplitude is given by
\begin{equation}
\langle 1|e^{-\,iHt}|1 \rangle = 
\frac{1}{2\pi i}\,\int_{-\infty }^{\infty }\,d\omega \,
\tilde{\Delta } (\omega + i0^{+})e^{-i\omega t} \,.
\end{equation}
Thus, there exist two Fourier decompositions for the same object, this one
and the one in the text. These two must be equivalent;
\( \:
F(\omega + i0^{+}) = \tilde{\Delta }(\omega + i0^{+}) \,.
\: \)
This further indicates the equality of the spectral functions;
$\rho (\omega ) = H(\omega )$.
We shall show how this spectral function is related to
the usual Lehmann spectral function $\rho _{L}(\omega ^{2})$.

If one takes the initial state to be an eigenstate of unperturbative
Hamiltonian $H_{0}$ (this automatically holds in the projector formalism), 
then one has a well-known relation to the time evolution operator;
\begin{equation}
\langle 1|e^{-\,iHt}|1 \rangle = e^{-iE_{1}t}\,\langle 1|U(t \,, 0)| 1 \rangle
= e^{-iE_{1}t}\,c(t) \,, \hspace{0.5cm} 
U(t \,, 0) = e^{iH_{0}t}\,e^{-\,iHt} \,.
\end{equation}
As a physical requirement we impose that an unstable particle state
$|1\rangle $ has no overlap with the true vacuum.
This suggests to take as the initial state one particle state of the form,
\begin{equation}
|1\rangle = -\,i\,\int\,d^{3}x\,f_{0}
(\vec{x} \,, t_{i})\stackrel{\leftrightarrow }
{\partial }_{0}\varphi (\vec{x} \,, t_{i})\,|0\rangle \,, 
\end{equation}
since $\langle 0|\varphi (x)|0 \rangle = 0$.
Here $f_{0}(\vec{x} \,, t_{i})$ is the mode function appropriate
for the initial state, $\varphi (\vec{x} \,, t_{i})$ being the Heisenberg
field operator, and $|0\rangle$ is the true vacuum of the total Hamiltonian.
The non-decay amplitude is thus written in terms of the Green's function,
\( \:
\langle 0|\varphi (t)\,\varphi (t_{i})|0 \rangle \,;
\: \)
\begin{eqnarray}
&& \hspace*{-1cm}
c(t) = e^{iM(t - t_{i})}\,
\frac{(i\partial _{t} + M)^{2} }{2M}\,
\int\,d^{3}x\,\int\,d^{3}x_{i}\,
\langle 0|T(\varphi (x)\varphi (x_{i}))|0 \rangle
\,,
\\ && 
\Delta (z^{2}) = \int_{\omega _{c}^{2}}^{\infty }\,d\sigma ^{2}\,
\frac{\rho _{L}(\sigma ^{2})}{z^{2} - \sigma ^{2}} \,.
\end{eqnarray}
The Lehmann representation for the renormalized Green's function
leads to
\begin{eqnarray}
\langle 1|e^{-iH(t - t_{i})}|1 \rangle &=&
\frac{1}{2\pi i}\,
\frac{(i\partial _{t} + M)^{2} }{2M}\,
\int_{-\infty }^{\infty }\,d\omega \,
e^{-i\omega (t - t_{i})}\, \Delta (\omega ^{2} + i0^{+}) \,.
\end{eqnarray}
If the spectral function is well convergent at $\omega \rightarrow \infty $,
\begin{equation}
c(t) = e^{iM(t - t_{i})}\,
\int_{\omega _{c}}^{\infty }\,d\omega \,\frac{(\omega + M)^{2}}{2M}\,
\rho _{L}(\omega ^{2}) \,e^{-i\omega (t - t_{i})} 
\,.
\end{equation}
The two spectral functions are then related by
\begin{equation}
\rho (\omega ) = \frac{(\omega + M)^{2}}{2M}\,\rho _{L}(\omega ^{2}) \,.
\end{equation}

Presence of the factor $(\omega + M)^{2}/2M$ in the last form of
the non-decay amplitude casts a doubt on validity of 
exchanging the order of derivative and integration.
There are however cases in which it is possible to introduce a regulator
and deform the contour of $\omega $ integration such that
the resultant renormalized formula
is convergent even after taking the infinite regulator mass limit.

The rest of arguments depends on existence of the asymptotic field 
corresponding to the unstable particle and 
perturbative expansion of the proper self-energy $G(\omega ^{2})$,
\begin{equation}
\Delta (\omega ^{2})^{-1} = \omega ^{2} - M^{2} -
G(\omega ^{2}) \,. 
\end{equation}
The proper self-energy $G(\omega ^{2})$ can be made finite by
using the renormalized perturbation scheme (with necessary couter terms
included as interaction terms).
In effect, this renormalization is equivalent to changing the coefficient
of $\omega ^{2}$ term in $\Delta (\omega ^{2})^{-1}$, to be later fixed by
the normalization condition of the renormalized initial amplitude,
\( \:
c_{R}(0) = 1 \,,
\: \)
and at the same time using a twice-subtracted renormalized
\( \:
G_{R}(\omega ^{2})
\: \)
instead of $G(\omega ^{2})$;
\begin{eqnarray}
&& \hspace*{-2cm}
e^{-iM(t - t_{i})}\,c_{R}(t) = 
\frac{1}{2\pi i}\,
\frac{(i\partial _{t} + M)^{2} }{2M}\,
\int_{-\infty }^{\infty }\,d\omega \,
e^{-i\omega (t - t_{i})}\,
\frac{1}{N\,\omega ^{2} - M^{2} - G_{R}(\omega ^{2})} 
\,, \label{green function form} 
\\ &&
G_{R}(\omega ^{2}) =
2\,\omega ^{4}\,\int_{\omega _{c}}^{\infty }\,d\omega '\,
\frac{2M\,\tilde{\sigma }(\omega ')}
{\omega '\,^{3}\,(\omega ^{2} - \omega '\,^{2} + i0^{+}\,)} \,,
\end{eqnarray}
with $M$ here the renormalized mass.
The spectral function $\tilde{\sigma }$
here coincides with $\sigma (\omega )$ of the
projector formalism to lowest order of perturbation.
In lowest order of perturbation one may use $\sigma _{i}(\omega )$ 
already given for various models in the text.

We note that eq.(\ref{green function form}) is not well defined
as an integral on the real axis,
if the time derivative operation is performed within the $\omega $ integral,
due to a bad high energy behavior even in renormalizable models;
\( \:
\sigma (\omega ) \propto \omega ^{2}
\: \)
as $\omega \rightarrow \infty $.
The perturbative infinity mentioned in \cite{maiani et al 93} 
is related to this ill defined formula (actually its expansion
in terms of the proper self-energy).

For the renormalized amplitude $c_{R}(t)$ it is more convenient 
to deform the integration contour from the real axis.
To any finite orders of perturbation there are only a finite number
of branch cut singularities in the complex $\omega $ plane and
deformation is possible passing through a finite number of Riemann sheets.
Thus one may deform the contour into the lower half plane of $\Re \omega >0$,
where $e^{-i\omega t}$ gives a good convergence factor.
In lowest order of perturbation one may use
the complex integral formula encircling both the pole (C$_{0}$) and
C$_{1}$ in Fig.1. Both contributions 
are well convergent even in the infinite regulator limit.
The formula thus defined makes it posssible to estimate the 
short-time limit of the non-decay amplitude.

We shall discuss the renormalizable model of fermion-pair decay.
An explicit calculation in this model gives
\begin{eqnarray}
G_{R}(\omega ^{2}) &=& -\,\frac{g^{2}}{4\pi ^{2}}\,
\left( \, 4m^{2} - \frac{4}{3}\omega ^{2} - 
\frac{(\omega ^{2} - 4m^{2})^{3/2}}{\omega }\,
\ln \frac{\omega - \sqrt{\omega ^{2} - 4m^{2}}}{2m} \,\right)
\nonumber 
\\ 
&+& i\frac{g^{2}}{8\pi }\,\frac{(\omega ^{2} - 4m^{2})^{3/2}}{\omega }
\,.
\end{eqnarray}
The C$_{1}$ integral, in the $t \rightarrow 0$ limit but for
$Mt \gg e^{- 4\pi^{2} /g^{2}}$, gives
\begin{equation}
\langle 1|e^{-iHt}|1 \rangle \approx 
-\,iN\frac{g^{2}}{16\pi ^{2}\,M}\,e^{-i\omega _{c}t}\,\int_{0}^{\infty }\,
dy\,e^{- yt} =
-\,iN\frac{g^{2}}{16\pi ^{2}\,Mt}\,e^{-i\omega _{c}t} \,,
\label{short-time limit} 
\end{equation}
where $N$ is a constant to be determined by the wave function 
renormalization.
The linear rise ($\propto 1/t$) found here is surprising, but is
inevitable if one can ignore the non-perturbative region 
of $\omega $ integral,
\( \:
\omega > M\,e^{4\pi^{2} /g^{2}} \,.
\: \)
The transition time from this very early to the exponential period
may be estimated by equating the above amplitude to the exponential
formula. The result is
\begin{equation}
t \approx \frac{\Gamma }{2\pi \,M^{2}} \,.
\end{equation}
This time scale is even much shorter than $1/M$, and is beyond
the reach of experimental search.
A numerical evidence supporting the linear rise of the short-time limit
is shown in Fig.8.

In the super-renormalizable model of boson-pair decay
the short-time limit is different from the one in the renormalizable
model above. It gives the linear decrease
of $1- O[t]$ for the non-decay amplitude, 
however with a constant of the linear term different from $\Gamma $
expected from the exponential law.

%\newpage

\newpage
\begin{center}
\begin{Large}
{\bf Figure Caption}
\end{Large}
\end{center}

\vspace{1cm} 
{\bf Fig.1}\\
Contours for integration of the non-decay amplitude.
Dashed is the part of contour in the second Riemann sheet, with the dashed
cross the location of the pole.

\vspace{0.5cm} 
{\bf Fig.2}\\
Comparison of large-time formulas. The solid curve is 
the real part of the non-decay amplitude obtained from an
exact numerical integration along the contour $C_{1}$ as in Fig.1,
and the dashed one is the approximate formula in the text,
the magnitude of amplitude computed in an arbitrary unit.
A phase factor is taken out such that the real part of the exact amplitude
becomes real and positive as $t\rightarrow \infty $. At times $< 8.5$
the exact real part becomes negative. 

\vspace{0.5cm} 
{\bf Fig.3}\\
Three types of diagrams.

\vspace{0.5cm} 
{\bf Fig.4}\\
Non-decay probability for the fermion-pair model.
In the inlet is shown an enlarged short-time behavior.
Chosen parameters, both for this and figure 5, are
\( \:
g = \frac{1}{3}\,, \hspace{0.5cm} Q \sim 0.8977 \,M_{0}\,, \hspace{0.5cm} 
M \sim 0.9977 \,M_{0}\,,
\: \)
with $M_{0}$ the bare parent mass.
A Lorentzian resolution function with $M_{0}\,\Delta t = 20$ is used.

\vspace{0.5cm} 
{\bf Fig.5}\\
Comparison of the short-time behavior with its limiting formula.
The dashed curve is the pole approximation whose parameters are
numerically obtained, while the dash-dotted one is the short-time
limit in the text, eq(\ref{short-time formula}).

\vspace{0.5cm} 
{\bf Fig.6}\\
Pole location in the second Riemann sheet for the S-wave (crossed), and
the P-wave(black circle) decays.
Different $Q$ values are computed
for a fixed $g = 1.0$ and $\Delta t = 20$ in the unit of the bare
mass $M_{0} = 1$.
The dotted and the dashed curves are a fit of the $\Gamma - Q$ relation
in perturbation theory.
The S-wave curve is better fitted with a formula,
$\Gamma = \alpha \sqrt{Q + \beta }$, with  constant $\alpha \,, \beta $.
The point marked by the crossed box is the parameter set for
Fig.7.

\vspace{0.5cm} 
{\bf Fig.7}\\
S-wave non-decay probability.
Chosen parameters for the boson-pair decay model are
\( \:
\mu  = 1.0\times M_{0} 
\,, \hspace{0.5cm}  M \sim 0.9988\,M_{0} \,, \hspace{0.5cm} 
Q \sim 1.07\times 10^{-5} \,M_{0}\,.
\: \)
The perturbative decay rate $\Gamma \approx 1\times 10^{-4}\,M_{0}$.
A Lorentzian resolution function with $M_{0}\,\Delta t = 20$ is used.

\vspace{0.5cm} 
{\bf Fig.8}\\
Short-time limit in the field theoretical approach outlined in Appendix.
An exact numerical computation of C$_{1}$ integral (the absolute value
by the solid line and a dominant real amplitude by the dash-dotted line)
is compared to eq.(\ref{short-time limit}) (the broken line) supporting
$1/t$ rise in the fermion-pair decay model.
Chosen parameters are the same as in Fig.2.

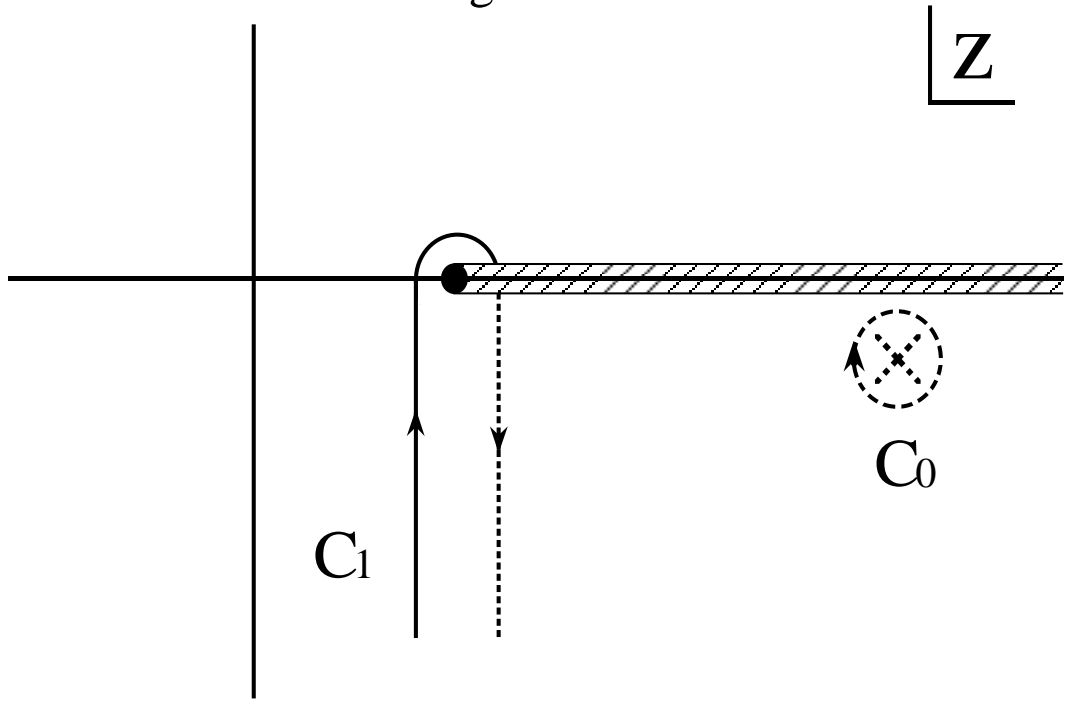
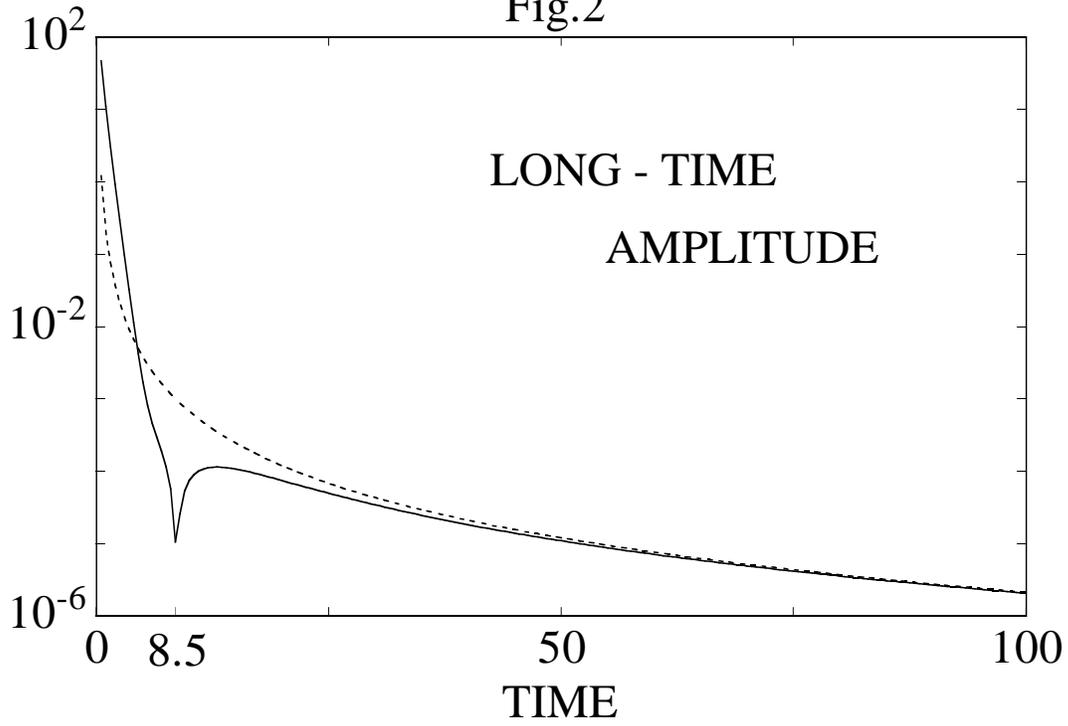
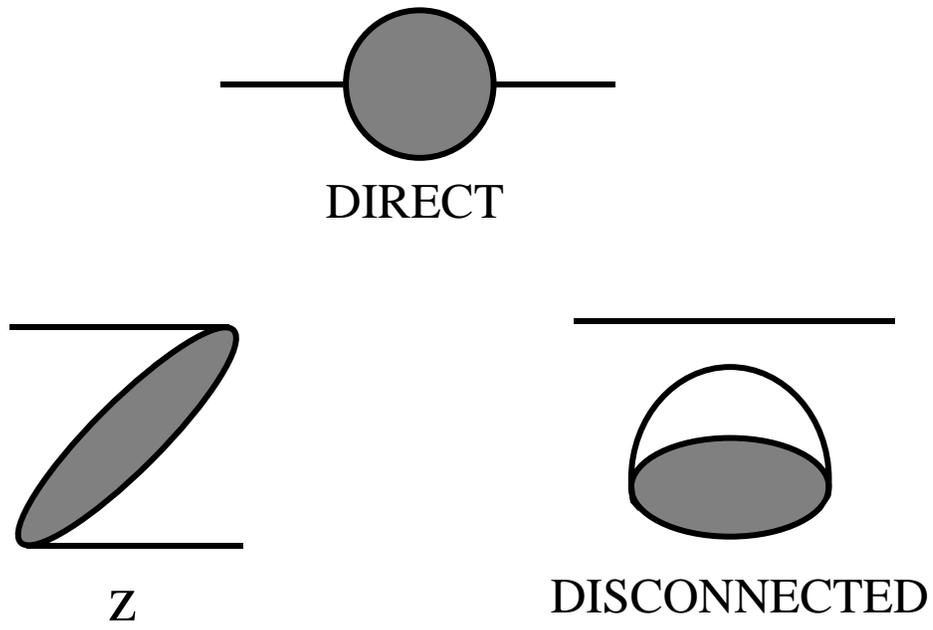
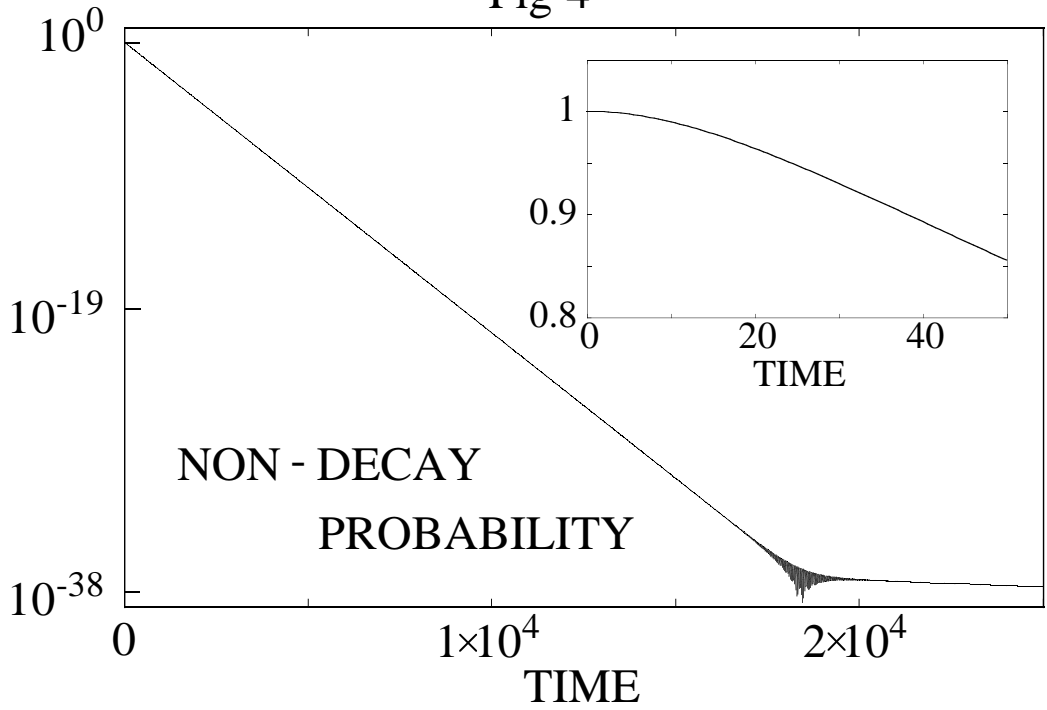
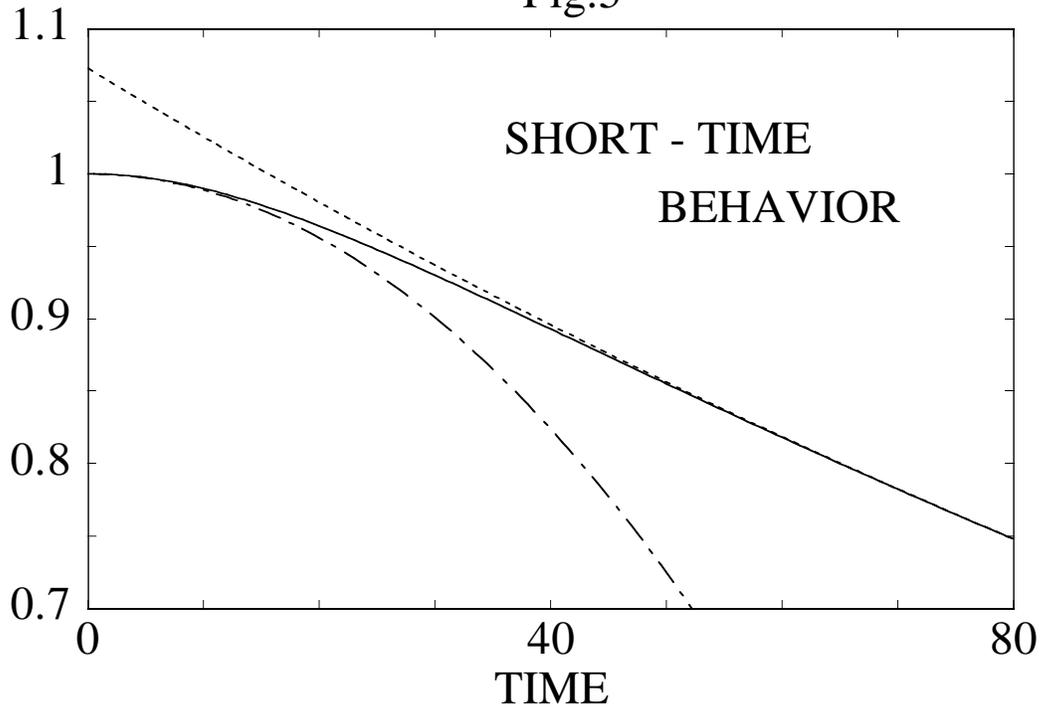
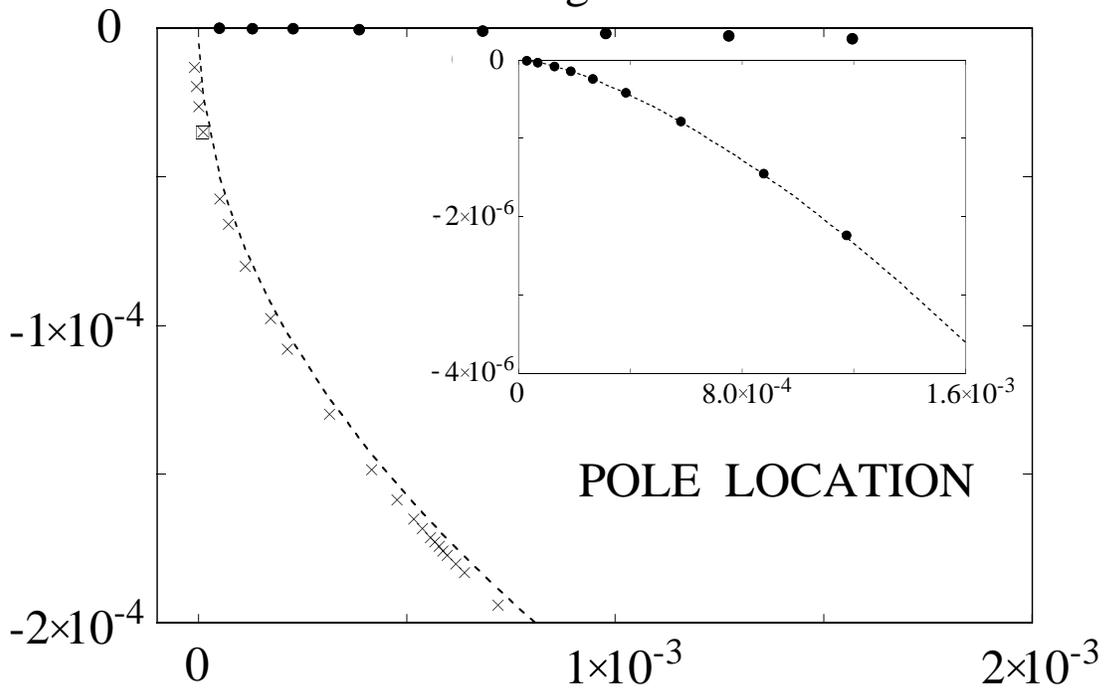
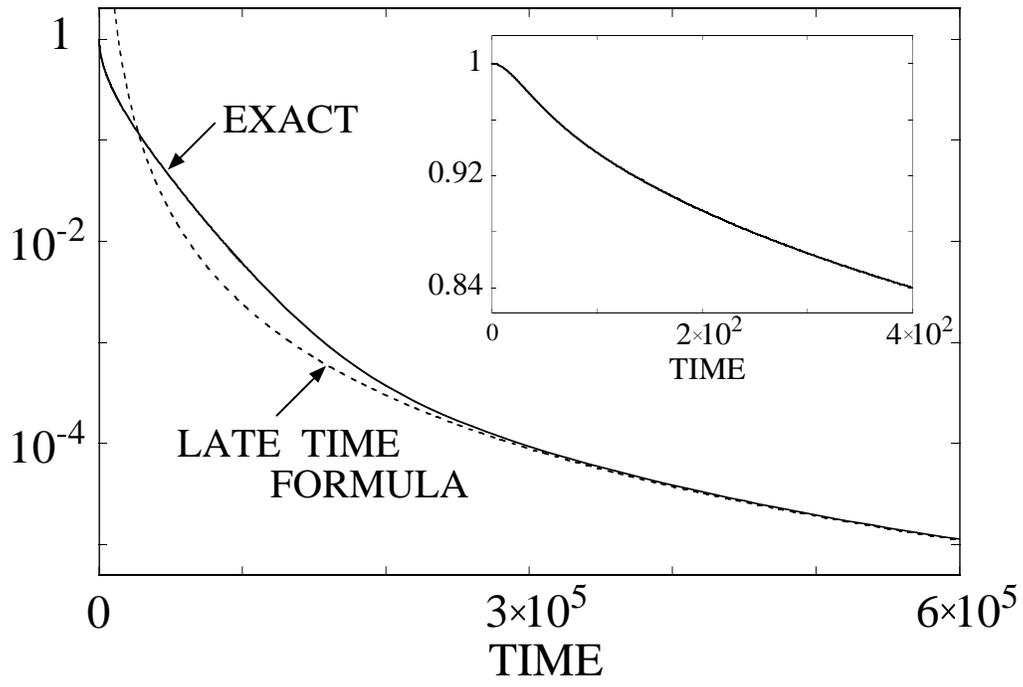
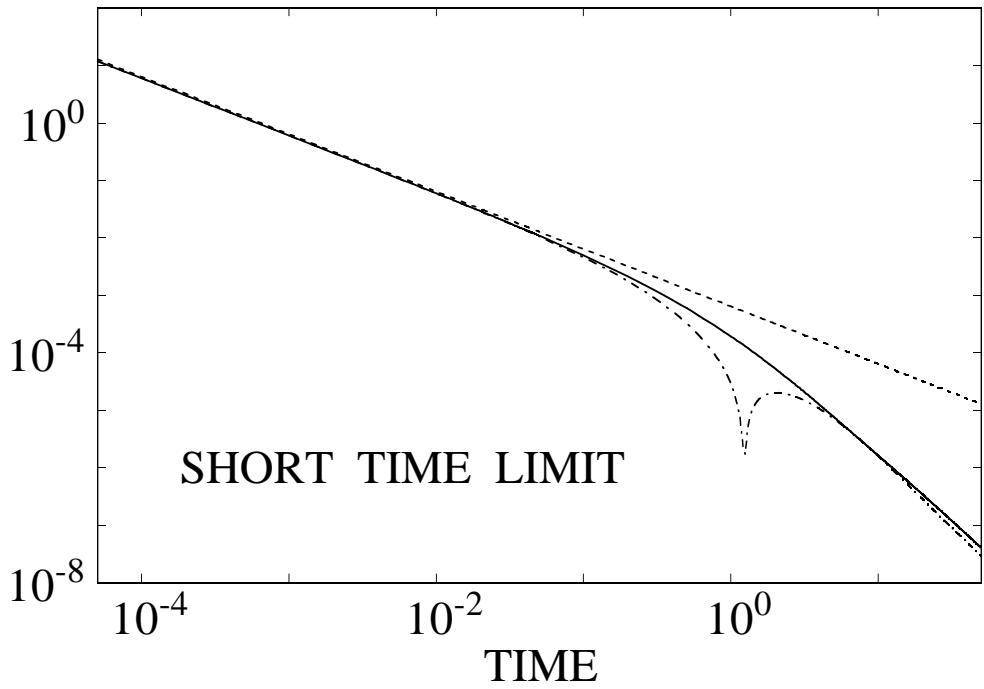
\end{document}